\documentclass[showpacs,preprintnumbers,preprint]{revtex4}
\usepackage{amsmath}
\usepackage{graphicx}
\usepackage{dcolumn}
\usepackage{bm}
\usepackage{amssymb}
\usepackage{makeidx}
\usepackage{amsmath}
\usepackage{lscape}
\usepackage{epsfig}
\newcommand{\ri}{{\mathrm i}}
\newcommand{\p}{\partial}
\newcommand{\bea}{\begin{array}}
\newcommand{\eea}{\end{array}}

\catcode`@=11 \long
\def\@caption#1[#2]#3{\par\addcontentsline{\csname
ext@#1\endcsname}{#1} {\protect\numberline{\csname
the#1\endcsname}{\ignorespaces #2}} \begingroup \small
\@parboxrestore \@makecaption{\csname fnum@#1\endcsname}
{\ignorespaces #3}\par \endgroup} \catcode`@=12

\newcommand{\la}{\label}

\catcode`@=11 \long
\def\@caption#1[#2]#3{\par\addcontentsline{\csname
ext@#1\endcsname}{#1} {\protect\numberline{\csname
the#1\endcsname}{\ignorespaces #2}} \begingroup \small
\@parboxrestore \@makecaption{\csname fnum@#1\endcsname}
{\ignorespaces #3}\par \endgroup} \catcode`@=12

\begin{document}

\allowdisplaybreaks
  \title {  Laplace-Runge-Lenz vector for arbitrary spin}
\author{A.G. Nikitin}
\email{nikitin@imath.kiev.ua} \affiliation{ Institute of
Mathematics, National Academy of Sciences of Ukraine,\\ 3
Tereshchenkivs'ka Street, Kyiv-4, Ukraine, 01601}
 \date{\today}

\begin{abstract}
A countable set of superintegrable quantum mechanical systems is
presented which admit the dynamical symmetry with respect to algebra
so(4). This algebra is generated by the Laplace-Runge-Lenz vector
generalized to the case of arbitrary spin. The presented systems
describe neutral particles with non-trivial multipole momenta. Their
spectra can be found algebraically like in the case of Hydrogen
atom. Solutions for the systems with spins 1/2 and 1 are presented
explicitly, solutions for spin 3/2 are expressed via solutions of an
ordinary differential equation of first order.
\end{abstract}
\maketitle
\section{Introduction\label{intro}}
The Laplace-Runge-Lenz (LRL) vector is a fundamental constant of motion admitted by the classical Kepler problem. That is a corner stone of celestial mechanics.  It also plays a very important role in quantum mechanics, starting with its  impressive  application by Pauli \cite{PAULI} who had found the spectrum of the Hydrogen atom before the Schr\"odinger equation was discovered.

However, Pauli did not identify the related symmetry group. It was
done later by Fock \cite{Fock} who discovered the  symmetry of the
Schr\"odinger equation for the Hydrogen atom w.r.t. group SO(4) for
bound states and group SO(1,3) for states with continuous spectrum.
Then Bargman \cite{Bargman} showed that Fock's group is generated by
the integral of motion which is a  quantum analogue of LRL vector.

  Since the Hydrogen atom admits six integrals of motion including  LRL and the orbital momentum vector, it is a maximally superintegrable system.  The number of its algebraically independent integrals of motion is equal to 5, and three of them (including the Hamiltonian) commute each other.   Such systems have many attractive properties starting with multi separability (which is guaranteed if integrals of motion are  second order differential operators) and ending with simple solutions which are polynomials multiplied by some overall factor.  Some of that systems, e.g.,  Hydrogen atom and harmonic oscillator are exceptionally important in physical applications.

   Systematic search for  superintegrable systems is a very interesting and
   important business started some time ago in papers \cite{wint1} and
   \cite{wint2}. A relatively new field is a classification of
   superintegrable systems with spin, which include matrix potentials.  The systems with spin-orbit coupling
   where discussed in recent papers \cite{wint4}-\cite{wint3} while the
   systems with Pauli interaction where investigated in \cite{N5}-\cite{N3}.
Let me mention also earlier papers \cite{bec1} and \cite{bec2} where
higher order integrals of motion of quantum mechanical systems with
$2\times2$ and $3\times3$ matrix potentials where investigated.

   A specific sector of the superintegrability phenomena
   is formed by QM systems which admit a (generalized) LRL vector.
There are various generalizations of the LRL vector in classical
  mechanics, see, e.g. \cite{gen1}-\cite{Yak}. However,  we can find only
  few examples of 3d quantum mechanical systems with spin which admit this
  symmetry. Namely, they are the MICZ-Kepler system which includes
   a magnetic monopole  with an arbitrary electric
charge and a fixed inverse-square term in the potential \cite{Zvan},
\cite{McI}, a charged particle with   spin $\frac12$
 and gyromagnetic ratio g = 4, interacting with a superposition of a point
magnetic monopole field and a Coulomb plus a fine-tuned
inverse-square potential \cite{Dyon1}, \cite{Dyon2}, a system with
spin-orbit interaction and a special inverse-square potential with
{\it fixed} coupling constant \cite{wint3}, and neutral particle
with spin $\frac12$ and a non-trivial dipole momentum interacting
with the external field inverse in radius \cite{N1}. The latter
system is shape invariant and was solved using tools of SUSY quantum
mechanics.

 Since $s=\frac12$ is only a particular (although very important) possible
 spin value, a natural desire is to generalize the results of \cite{N1} to
 the case of arbitrary spin $s$. This problem is seemed to be especially
 provocative  since in the 2d case the systems admitting the generalized
 LRL vector where found for both spin $s=\frac12$ \cite{pron1} and  $s$
 arbitrary \cite{pron2}, \cite{N2}.

In the present paper the 3d LRL vectors for arbitrary spin are
introduced. Namely, a countable set of exactly solvable QM problems
is presented which admit generalized LRL vectors with spin. The
geometric symmetry of these problems is exhausted by their
invariance w.r.t. the rotation group generated by the total orbital
momentum  including a spin vector with arbitrary $s$. However, any
of the discussed problems admits three additional constants of
motion which are components of the LRL vector and generate the
dynamical symmetry w.r.t. group SO(4).

This extended symmetry makes it possible to find algebraically Hamiltonian eigenvalues and essentially simplifies the procedure of construction of the corresponding eigenvectors. The Hamiltonian eigenvectors for spins $\frac12$ and $1$ are presented in explicit form, those ones for spin $\frac32$ are expressed via solutions of an ordinary differential equation of the fourth  order.

The presented systems describe neutral particles with (arbitrary)
spin, which have non-trivial dipole or multipole momenta and are
affected by dipole or multipole interactions with the external
fields.

\section{Runge-Lenz vector for Hydrogen atom}

Let us start with the well known example of the dynamical symmetry
w.r.t. group SO(4), which is admitted by the hydrogen atom. This QM
system is specified by the following hamiltonian:
\begin{gather}H=\frac{p^2}{2m}+V(x)\la{H1}\end{gather}
where
\[p^2=p_1^2+p_2^2+p_3^2,\quad p_1=-i\frac{\p}{\p x_1},\quad V=-\frac{\alpha}{x},\quad
x=\sqrt{x_1^2+x_2^2+x_3^2},\ \alpha=e^2>0.\]

We will discuss the eigenvalue problem for hamiltonian (\ref{H1})
\begin{gather}\la{ep1} H\psi=E\psi\end{gather}
and recall  an elegant way to find eigenvalues $E$ algebraically,
without solving the stationary Schr\"odinger equation (\ref{ep1}).

Hamiltonian (\ref{H1})  commutes with generators of group O(3) which
are components of the orbital momentum ${\bf L}={\bf x}\times{\bf
p}.$ We can indicate  three additional constants of motion which are
components  of the LRL vector:
\begin{gather}\la{RuL}{\bf K}=\frac1{2m}({\bf p}\times {\bf L}-
{\bf L}\times{\bf p})+{\bf x}V. \end{gather}
The components of $\bf L$ and $\bf K$ satisfy the following commutation relations:
\begin{gather}\la{co}[L_a,H]=[K_a,H]=0,\\\la{com}\begin{split}&[L_a,L_b]
=\ri\varepsilon_{abc}L_c,\quad
[K_a,L_b]=\ri\varepsilon_{abc}K_c,\\&[K_a,K_b]=-\frac{2\ri}{m}
\varepsilon_{abc}L_c
H.
\end{split}\end{gather}

Changing $H$ in (\ref{co}) by its eigenvalue $E$ we obtain the Lie
algebra isomorphic to so(4) provided $E<0$ or to o(1,3) if $E$ is
positive.

It is the hidden symmetry w.r.t. group so(4) which causes the
degeneration of the spectrum of hamiltonian (\ref{H1}) w.r.t. the
orbital quantum number ${ l}$. This symmetry also causes the maximal
superintegrability of the Hydrogen atom. Moreover, we can find the
eigenvalues $E$ of hamiltonian (\ref{H1}) algebraically. Let us
present this well known procedure which will be used in the
following for another models.

First we rescale $\bf K$ and consider the following vectors:
\begin{gather}\la{so4}
{\bf K}'=\sqrt{-\frac{m}{2E}}{\bf K},\quad {\bf g}=\frac12({\bf L}+{\bf K}'),\quad {\bf q}=\frac12({\bf L}-{\bf K}').\end{gather} Their components satisfy the following commutation relations
\begin{gather}\la{su2}[g_a,g_b]=i\varepsilon_{abc}g_c,\quad [q_a,q_b]=i\varepsilon_{abc}q_c,\quad [g_a,q_b]=0\end{gather} and so form a basis of the Lie algebra so(4)$\cong$su(2)$\bigoplus$su(2). The related Casimir operators are
\begin{gather}\la{Cas0}C_-=4{\bf q}^2,\quad C_+=4{\bf g}^2.\end{gather}

On the other hand, in accordance with (\ref{so4}),
\begin{gather}\la{ev}C_\pm={\bf L}^2+\nu^2{\bf K}^2\pm 2\nu{\bf L}\cdot{\bf K}\end{gather} where $\nu=\sqrt{-\frac{m}{2E}}$.
 Since ${\bf L}\cdot{\bf K}\equiv0$, in our case $C_+=C_-=4{\bf q}^2=4{\bf g}^2$.

In accordance with (\ref{su2}) components $q_a$ of vector $\bf q$ form a basis of algebra su(2). Its irreducible representations are labeled by integers of half integers $q$. The same is true for vector $\bf g$. The corresponding eigenvalues $c_\pm$ of the Casimir operator $C_\pm$ are:
\begin{gather}\la{c1} c_-=4q(q+1),\quad c_+=4g(g+1)\end{gather}
where $q$ and $g$ are non-negative integers or half integers.

As it follows from definition (\ref{RuL})
\begin{gather}\la{K2}{\bf K}^2=\alpha^2+({\bf L}^2+1)\frac{2H}m\end{gather} and so equation (\ref{ev}) gives: $C_1=-1-\frac{\alpha^2m}{2E}.$ Comparing that with
(\ref{c1}) one obtains
\begin{gather}\la{E}E=-\frac{m\alpha^2}{2n^2},\quad n=2q+1=1,2,...\end{gather}

Thus, using the commutativity of hamiltonian (\ref{H1}) with LRL vector, it is possible to obtain the  energy levels of Hydrogen atom algebraically.

\section{Superintegrable systems with arbitrary spin}
The model hamiltonian (\ref{H1}) ignores the spin of the orbital
electron which  in fact is treated as a scalar particle. To
introduce a spin it is necessary to change the angular momentum $\bf
L$ by the total angular momentum
\begin{gather}\la{tm}{\bf L}\to{\bf J}={\bf L}+{\bf S}\end{gather}
where ${\bf S}$ is the spin vector whose components are matrices satisfying the following relations: \begin{gather}\la{spin}[S_a,S_b]=\ri\varepsilon_{abc}S_c,\quad S_1^2+S_2^2+S_3^2=s(s+1)I\end{gather} where $I$ is the unit matrix.

To obtain  the corresponding Runge-Lenz vector we change in (\ref{RuL}) ${\bf L}\to{\bf J}$ and  $V\to\hat V$ where $\hat V$ is an unknown potential. As a result we obtain:
\begin{gather}\la{GRL}{\hat {\bf K}}=\frac1{2m}({\bf p}\times {\bf J}-
{\bf J}\times {\bf p})+{\bf x}\hat V.\end{gather}

By definition vector (\ref{GRL}) should commute with the hamiltonian
\begin{gather}\la{HS}\hat H=\frac{p^2}{2m} +\hat V.\end{gather} It is the case iff potential $\hat V$ satisfies the following conditions:
\begin{gather}[\hat V,\bf J]=0,\la{cond1}\\
{\bf x}\cdot\nabla \hat V+\hat V=0,\la{cond2}\\
{\bf S}\times(\nabla\hat V)-(\nabla\hat V)\times{\bf
S}=0\la{cond3}\end{gather} where $\cdot$ and $\times$ are the
symbols of the scalar and vector products and $\nabla$ is the
gradient vector.

If conditions (\ref{cond1}) -- (\ref{cond3}) are satisfied, then
operators $\bf J$, $\hat {\bf K}$ and $\hat H$ satisfy the same
relations as the orbital momentum, LRL vector and hamiltonian of the
Hydrogen atom, i.e.,
 \begin{gather}\la{coco}[J_a,H]=[{\hat K}_a, H]=0,\\\la{como}\begin{split}&[J_a,J_b]
=\ri\varepsilon_{abc}J_c,\quad [{\hat K}_a, J_b]=\ri\varepsilon_{abc}{\hat K}_c,
\\& [{\hat K}_a,{\hat K}_b]=-\frac{2\ri}{m}\varepsilon_{abc}J_c
H.
\end{split}\end{gather}

 Let us calculate potential $\hat V$ for arbitrary spin. In accordance with
 (\ref{cond1}) $\hat V$ should be a scalar and so it can be expanded via the complete set of projection operators:
 \begin{gather}\la{VAS}\hat V=\sum_{\nu=-s}^sf_\nu\Lambda_\nu\end{gather}
where $f_\nu$ are functions of $r$ and $\Lambda_\nu$ are projectors onto the eigenvector space  of matrix
${\bf S}\cdot{\bf
n}, {\bf n}=\frac{\bf x}{x}$,  corresponding to the eigenvalue $\nu$ ($\nu,
\nu'=s,s-1,...-s$):
\begin{gather}\label{Lnu}\Lambda_\nu=\prod_{\nu'\neq\nu}
\frac{{\bf S}\cdot{\bf n}-\nu'}{\nu-\nu'}.\end{gather} Matrices
$\Lambda_\nu$ satisfy the standard projector properties
\[\Lambda_\nu\lambda_{\nu'}=\delta_{\nu\nu'}\Lambda_\nu,\quad
\sum_{\nu=-s}^s\Lambda_\nu=I,\quad {\bf S}\cdot{\bf
n}=\sum_{\nu=-s}^s\nu\Lambda_\nu\] and form a very convenient basis
for expansion of other scalar matrices.

 Operators (\ref{VAS})
satisfy condition (\ref{cond2}) iff functions $f_\nu(x)$ are
proportional to the inverse radius:
\begin{equation}\label{c_nu}
    f_\nu(x)=\frac{c_\nu}{x}
\end{equation}
where $c_\nu$ are constants.

Let us substitute (\ref{VAS}) and (\ref{c_nu}) into the remaining equation
(\ref{cond2}). To calculate the gradients of the projection operators we
use the following relations \cite{FGN}, \cite{FN}
\begin{gather}\label{com_re}
\nabla\Lambda_\nu=\frac{\ri}{2x}{\bf n}\times{\bf S}
(2\Lambda_\nu-\Lambda_{\nu+1}-\Lambda_{\nu-1})-\frac1{2x}\left({\bf
S}- {\bf n}({\bf S}\cdot{\bf n})\right)\left(\Lambda_{\nu+1}-
\Lambda_{\nu-1}\right).\end{gather}

 As a result, equating coefficients for linearly independent terms,
we obtain the following conditions for coefficients $c_\nu$:
\begin{equation}\label{nu_cnu}
   \nu c_\nu=(\nu+1)c_{\nu+1},\quad \nu=-s, -s+1, ..., s.
\end{equation}

If spin is integer then $\nu$ can take zero value, wile for half-integer spins all $\nu$ are nonzero. Thus the general solutions of equations (\ref{nu_cnu}) are:
\begin{equation}\label{gs1}
    c_0=\alpha,\quad c_\nu=0, \quad \text{if}\quad \nu\neq0
\end{equation}if spin is integer, and
\begin{equation}\label{gs2}
    c_\nu=\frac{\alpha}{\nu}
\end{equation}
for half integer spin. Here $\alpha$ is an arbitrary real parameter.

Substituting (\ref{c_nu}), (\ref{gs1}), (\ref{gs2}) into (\ref{VAS})
we obtain explicit forms of potentials for arbitrary spins:
\begin{equation}\label{Potsi}
   {\hat V}=\frac{\alpha}{x}\hat \Lambda=\frac{\alpha}{x}\Lambda_0\quad \text{for integer spins},
\end{equation}
\begin{equation}\label{potsh}
   \hat V=\frac{\alpha}{x}\hat\Lambda=\frac{\alpha}{x}\sum_{\nu=-s}^s\frac{1}{\nu}\Lambda_\nu \quad \text{for half integer spins}.
\end{equation}

Formulae (\ref{HS}), (\ref{Potsi}) and (\ref{potsh}) give the
general form of hamiltonians of neutral particles with arbitrary
spin $s$, which commute with the generalized LRL vector (\ref{GRL}).
Notice that matrices $\hat\Lambda$ solve the following equations:
\begin{gather*}\hat\Lambda{\bf S}\cdot{\bf n}=0\la{a1}\end{gather*}
for integer $s$ and
\begin{gather*}\hat\Lambda{\bf S}\cdot{\bf
n}=I\la{a22}\end{gather*} for $s$ half integer, where $I$ is the
unit matrix of dimension $(2s+1)\times(2s+1)$.
\section{Irreducibility conditions}
 By construction, operators (\ref{GRL}) where $\hat V$ are potentials
  (\ref{Potsi}) or (\ref{potsh}) satisfy commutation relations (\ref{coco})
  and (\ref{como}).
   Thus all obtained hamiltonians admit a hidden symmetry w.r.t.
   algebra so(4), which make it possible to find their spectra without
   solving the Schr\"odinger equation.

We will consider  eigenvalue problems for  hamiltonians (\ref{HS}) which include potentials (\ref{Potsi}) and (\ref{potsh}):
\begin{gather}\la{ep11}\hat H\psi\equiv\left(\frac{p^2}{2m} +\frac\alpha{x}\hat \Lambda\right)\psi=E\psi.\end{gather}
The eigenvectors $\psi=\psi({\bf x}) $ are supposed to be
normalizable and vanishing at ${\bf x}=0$.

It is convenient to rescale independent variables making the change
\begin{gather}\la{change}{\bf x}\to{\bf r}=\sqrt{-2mE}{\bf x
}.\end{gather}
As a result we transform (\ref{EP}) to the following form:
\begin{gather}\la{resc}(1-\Delta)\psi=\frac{2k}{r} \Lambda\psi\end{gather}
where $\Delta=\frac{\p^2}{\p r_1^2}+\frac{\p^2}{\p
r_2^2}+\frac{\p^2}{\p r_3^2},\ \tilde{\bf n}={\bf n}=\frac{{\bf
r}}{r},\ r=\sqrt{{r}_1^2+{r}_2^2+{r}_3^2},$ and
\begin{gather}\la{nota}k=\alpha\nu=\alpha\sqrt{\frac{m}{-2E}}.\end{gather}

In analogy with (\ref{so4}) we introduce two commuting vectors $\hat
{\bf q}$ and $\hat {\bf q}$ such that
\begin{gather}\la{qg}\bf J=\hat
{\bf q}+\hat {\bf g},\quad \nu\hat {\bf K}=\hat {\bf q}-\hat {\bf
g}.\end{gather}

 Consider now additional conditions which can be imposed on $\psi$ thanks to the hidden symmetry of equation (\ref{resc})
  w.r.t. algebra so(4).
 First we notice that the Casimir operators (\ref{Cas0})
 for arbitrary spin take the following form (compare with (\ref{ev})):
 \begin{gather}\la{cas3}C_\pm={\bf J}^2+\nu^2\hat{\bf K}^2\pm2\nu{\bf J}
 \cdot\hat{\bf K},\end{gather}
 moreover,
 \begin{gather}\la{cas4}{\bf J}\cdot\hat{\bf K}=
 \left\lbrace\begin{matrix}0\phantom{abcd}\qquad \text{ for integer }\ s,
 \\\alpha s I\qquad \text{ for half integer }\ s.\end{matrix}\right.\end{gather}
.

 For irreducible representations of algebra so(4)
 the Casimir operators (\ref{cas3}) should be proportional to the unit
 matrix, and their eigenvalues are given by equation (\ref{c1}).
 Thus, in addition to the Schr\"odinger equation (\ref{ep11}) with
 Hamiltonian (\ref{HS}), we can impose two additional  constrains  on the
 wave function $\psi$:
 \begin{gather}\la{cas5}C_-\psi=4q(q+1)\psi,\quad
 C_+\psi=4g(g+1)\psi\end{gather}
where $C_-$ and $C_+$ are given by equations (\ref{cas3}),  $q$ and
$g$ are arbitrary integers. Substituting (\ref{cas3}) and
(\ref{tm}), (\ref{GRL}) into (\ref{cas5}) and  using equation
(\ref{resc}) we obtain the following systems:
\begin{gather}\la{case}\left(({\bf S}\cdot\tilde{\bf\nabla})^2+\frac{k}{r}({\bf S}\cdot{\bf J}\Lambda+\Lambda{\bf S}\cdot{\bf J})+k^2\Lambda^2\right)\psi=\omega_\pm\psi\end{gather}
where $\tilde{\bf \nabla}$ is the gradient vector for rescaled variables $\tilde{\bf r}$,
\begin{gather}\la{cas6}\omega_+=\omega_-=(2g+1)^2\end{gather} for integer $s$, and
\begin{gather}\la{cas7}\omega_+=(2g+1)^2-2sk,\quad \omega_-=(2q+1)^2+2sk\end{gather}
for $s$ half integer.

 In the following sections particular cases of Hamiltonians $\hat H$
 corresponding to $s=\frac12, 1$ and $\frac 32$ are considered in more
 detail. We will see that the compatibility condition for equations
 (\ref{resc}) and (\ref{case}) fixes the hamiltonian spectrum.
 \section{Laplace-Runge-Lenz vector for spin $\frac12$}

For the case $s=\frac12$ we have ${\bf S}=\frac12{\mbox{\boldmath
$\sigma$}}$,  components of vector ${\mbox{\boldmath $\sigma$}}$ are
Pauli matrices. The corresponding potential (\ref{potsh}) is reduced
to the following form:
\begin{gather}\la{V}\hat V=
\alpha\frac{\mbox{\boldmath $\sigma$} \cdot \bf x}{x^2}\end{gather}
while the related total orbital momentum (\ref{tm}) and LRL vector
(\ref{GRL}) can be written as:
\begin{gather}\la{N1}\begin{split}&{\bf J}={\bf L}+\frac12{\mbox{\boldmath
$\sigma$}},\\&{\hat {\bf K}}=\frac1{2m}({\bf p}\times {\bf J}- {\bf
J}\times {\bf p})+\alpha{\bf x}\frac{\mbox{\boldmath $\sigma$} \cdot
\bf x}{x^2}.\end{split}\end{gather}

In analogy with section 2 it is possible to find eigenvalues of
hamiltonian (\ref{HS}), (\ref{V}) algebraically, without solving the
corresponding  equation (\ref{ep11}). In accordance with (\ref{N1}),
 the square of the generalized LRL vector takes the
following form:
\begin{gather*}{\bf \hat K}^2=\left(2{\bf J}^2+\frac32\right)\frac{H}{m}+\alpha^2. \end{gather*}
Substituting  this expression into equation (\ref{cas3}) and setting $H=E$ we obtain:
\begin{gather}\la{casc}C_\pm=\left(\nu^2\alpha^2\pm\nu\alpha-
\frac34\right)I\quad \end{gather}
where $\nu=\sqrt{-\frac{m}{2E}}$.

It follows from (\ref{casc}) that for $s=\frac12$ equation
(\ref{case}) is reduced to an algebraic costraint. Indeed, in
irreducible representations of algebra su(2) $C_\pm=c_\pm I$ where
$c_\pm$ are given by equation (\ref{c1}). Thus we have algebraic
conditions
$$(\nu\alpha-1/2)^2=(2q+1)^2,\quad (\nu\alpha+1/2)^2=(2g+1)^2$$ for spectral parameter $\nu$, and so the eigenvalues $E$ for coupled states can be represented in the following form:
\begin{gather}\la{bb1} E=-\frac{m\alpha^2}{2N^2} \end{gather}where
\begin{gather}\la{bb11}N=n+1/2,\quad n=2q+1=2g=1, 2,\dots\end{gather}

Thus the energy spectrum for the system with spin $\frac12$ is given by
relation (\ref{bb1}) which is rather similar to the Balmer formula.
However, in contrast to the Hydrogen atom, the main quantum number $N$
should be half integer.

For a fixed $N$ the integrals of motion (\ref{N1}) generate an
irreducible representation $D({l}_0,{l}_1)$ of algebra so(4), where
\[{ l}_0=\frac12,\quad { l}_1=N\]
are the Gelfand-Tsetlin numbers which can expressed via $q$ and $g$
as $l_0=g-q,\ { {l}}_1=g+q+1$. This statement is a direct
consequence of equations (\ref{bb11}).

The spectrum (\ref{bb1}) and the corresponding eigenvectors were found in paper \cite{N1} using shape invariance of hamiltonian  (\ref{HS}), (\ref{V}). Let us  present a more straitforward way to obtain them.

Consider the eigenvalue problem for the hamiltonian including potential (\ref{V}):
\begin{gather}\la{bbb1}\left(\frac{p^2}{2m}+\alpha\frac{\mbox{\boldmath $\sigma$} \cdot \bf
x}{x^2}\right)\psi=E\psi. \end{gather}
Rescaling  independent variables in accordance with (\ref{change})
 we transform (\ref{bbb1}) to the following form:
\begin{gather}\la{resc1}(1+\tilde p^2)\psi=\frac{2k}{r} \Lambda\psi\end{gather}
where
$\Lambda={\mbox{\boldmath $\sigma$}} \cdot \tilde{\bf
n},\ \tilde{\bf n}={\bf n}=\frac{{\bf r}}{r}$ and $\tilde p=\sqrt{-2mE}p$.

Multiplying  (\ref{resc1}) by ${\mbox{\boldmath $\sigma$} \cdot \bf
r}$ we obtain:
\begin{gather}\la{1}{\mbox{\boldmath $\sigma$} \cdot \bf
r}(1+\tilde p^2)\psi=2k\psi.\end{gather}
Considering (\ref{1}) in the momentum representation we come to the system of the first order equations:
\begin{gather}\la{2}\ri{\mbox{\boldmath $\sigma$} \cdot \bf
\nabla}_{\tilde p}\Phi=\frac{2k}{1+\tilde p^2}\Phi\end{gather} where $\phi=(1+\tilde p^2)\psi$ and $\nabla_{\tilde p}$ is the gradient operator.

Introducing the spherical variables and expanding solutions of
equation (\ref{2}) via spherical harmonics:
 \begin{gather}\la{OOmu}\begin{split}&\Phi=\frac1p\sum_{j,\lambda,\kappa}
  \phi_{\lambda j\kappa}(p)
\Omega_{j,j-\lambda,\kappa}(\varphi,\theta),\quad \lambda=\pm\frac12,
\quad \kappa=-j,-j+1,...,j,\\
&\Omega_{j,j-\frac12,\kappa}=\begin{pmatrix}\sqrt{\frac{j+
\kappa}{2j}}Y_{j-\frac12,\kappa-\frac12}\\\sqrt{\frac{j-
\kappa}{2j}}Y_{j-\frac12,\kappa+\frac12}\end{pmatrix},\quad
\Omega_{j,j+\frac12,\kappa}=\begin{pmatrix}-\sqrt{\frac{j-
\kappa+1}{2j+2}}Y_{j+\frac12,\kappa-\frac12}\\\sqrt{\frac{j+
\kappa+1}{2j+2}}Y_{j+\frac12,\kappa+\frac12}\end{pmatrix}
\end{split}\end{gather}
where $ Y_{j\pm \frac12,
k\pm \frac12}$ are spherical functions,
 we come to the following system of radial equations:
\begin{gather*}\begin{split}&\ri\left(\frac\p{\p p}-\frac{2j+1}{2p}\right)\phi_{\frac12j\kappa}=\frac{2k}{1+p^2}\phi_{-\frac12j\kappa},\\&
\ri\left(\frac\p{\p p}+\frac{2j+1}{2p}\right)\phi_{-\frac12j\kappa}=\frac{2k}{1+p^2}\phi_{\frac12j\kappa}.
\end{split}\la{31}\end{gather*}
This system is solved by the following functions:
\begin{gather*}\begin{split}&\phi_{-\frac12j\kappa}=2k(j+1)p^{j+\frac32}
(1+p^2)^{-k}{\cal F}(-k+1,-k+j+1,j+2,-p^2),\\&\phi_{\frac12j\kappa}=
{\ri(p^2+1)^{-k}}\left(\left(
p^{j+\frac52}+p^{j+\frac92}\right)(k-1)(j+1-k){\cal F} (2-k, +2+j-k,
3+j, -p^2)\right.\\&\left.+(j+2)\left(j+1-k)
p^{j+\frac52}+(j+1)p^{j+\frac12}\right){\cal
F}(-k+1,-k+j+1,j+2,-p^2)\right)\end{split}\la{32}\end{gather*} where
${\cal F}(.,.,.,-p^2)$ are hypergeometric functions. These solutions
are square integrable provided $k=j+1+n, n=0,1,2,...$, which is in a
good accordance with equations (\ref{bb1}) and (\ref{nota}).

\section{The system with spin 1}
\subsection{Hamiltonian and other integrals of motion}
Let us consider the LRL vector for spin 1. It is defined by equations
(\ref{GRL}) and (\ref{tm}) where ${\bf S}=(S_1,S_2,S_3)$ is a matrix vector whose components can be chosen in  the following form:
\begin{gather}\la{mat}S_1=\frac1{\sqrt{2}}\left(\begin{matrix}0&1&0\\1&0&1\\
0&1&0\end{matrix}\right),\quad S_2=\frac{\ri}{\sqrt{2}}\left(\begin{matrix}0&-1&0\\
1&0&-1\\0&1&0\end{matrix}\right),\quad S_3=\left(\begin{matrix}1&0&0\\0&0&0\\
0&0&1\end{matrix}\right).\end{gather}
In accordance with (\ref{VAS}), (\ref{Lnu}) and (\ref{Potsi}) the corresponding potential in (\ref{GRL}) can be given by the following formula:
\begin{gather}\la{lala}
\hat V=\frac\alpha{x}(1-({\bf S}\cdot{\bf n})^2).\end{gather}
Thus the system with spin $s=1$ which admits the LRL vector is specified by the following hamiltonian
\begin{gather}\la{hamham}H=\frac{p^2}{2m} +
\frac\alpha{x}(1-({\bf S}\cdot{\bf n})^2).\end{gather}

Let us consider the eigenvalue problem for  hamiltonian (\ref{hamham}):
\begin{gather}\la{EP}\left(\frac{p^2}{2m}+\frac{\alpha}x(1-({\bf S}\cdot{\bf n})^2\right)\psi=E\psi.\end{gather}

Making the change (\ref{change}) we can transform (\ref{EP}) to the standard  form (\ref{resc}) where
\begin{gather}\la{rescaca}  \hat\Lambda=1-({\bf S}\cdot\tilde{\bf n})^2.\end{gather}

The corresponding system (\ref{resc}) includes three coupled equations  and so is rather complicated. Nevertheless it admits a sufficient number of integrals of motion to be integrated explicitly and in closed form.
These integrals of motion are the LRL vector and total angular momentum given by equations (\ref{GRL}), (\ref{tm}) where $\bf S$ and $\tilde V$ are matrices (\ref{mat}) and potential (\ref{lala}).

Thus we can
add to equation (\ref{EP}) the condition (\ref{case}).
 This system includes two arbitrary parameters, i.e., $k$ and $q$; the latter can take integer or half integer values. We will see that the compatibility condition for equations (\ref{EP}) and (\ref{case}) is nothing but an algebraic relation for these parameters, which defines possible values of $E$ in equation (\ref{EP}).

\subsection{Separation of variables and energy spectrum}
Let us start with equation (\ref{resc}). Exploiting its invariance
w.r.t. the rotation group whose generators are given by equations
(\ref{tm}) and (\ref{mat}) it is possible to separate variables. To
do it we introduce the spherical variables
 \begin{gather}\la{sv}x_1=x\cos\varphi\sin\theta,\ x_2=x\sin\varphi\sin\theta,\ x_3=x\cos\theta\end{gather} and expand $\psi=\psi({\bf x})$ via spherical harmonics $\Omega^s_{j,\kappa,\lambda}$ which are eigenvectors of the commuting integrals of motion $ J_3$, ${\bf J}^2
$ and ${\bf L}^2$ (see Appendix 1):
\begin{gather}\psi=\frac1x\sum_{j,\kappa,\ \lambda} \psi_{j, \kappa,\lambda}(x)
\Omega^s_{j,\kappa,\lambda}(\varphi,\theta).\la{bes}\end{gather} Here $s=1$,
$j=0,\ 1,\ \dots, \ \kappa=-j,\ -j+1,\ \dots\ j, \ \lambda=1, 0, -1$.

In the spherical harmonics basis, operator ${\bf L}^2$ and matrix
 ${\bf S}\cdot{\bf n}$  are reduced to the following matrices
 (see Appendix): ${\bf L}^2\to \hat { L}^2,\ {\bf S}\cdot{\bf n} \to
 \hat S_1$
 where
 \begin{gather}L^2= \left(\begin{matrix}j(j-1)&0&0\\0&j(j+1)&0\\0&0&(j+1)(j+2)
 \end{matrix}\right),\ \hat S_1=- \frac1{\sqrt{2j+1}}\left(\begin{matrix}0&\sqrt{j+1}&0\\\sqrt{j+1}&0&\sqrt{j}\\
 0&\sqrt{j}&0
 \end{matrix}\right).\la{s1}\end{gather}
 Substituting (\ref{bes}) and (\ref{s1}) into (\ref{EP}) and using variables (\ref{change}) we obtain the following equation for radial functions:
 \begin{gather}\la{sc}\frac{\p^2}{\p r^2}\Phi=\left(1+\frac{{\bf L}^2}{r^2}-
 \frac{2k}{r}\tilde\Lambda\right)\Phi\end{gather}
 where $\tilde\Lambda=1-{\hat S}_1^2$ and
 \begin{gather}\la{not1}\Phi=\text{column}\left(\Phi_-,\
 \Phi_{0},\ \Phi_{+}\right),\qquad \Phi_-=\psi_{j,\kappa,-1},\quad
 \Phi_0=\psi_{j,\kappa,0},\quad \Phi_+=\psi_{j,\kappa,1}.\end{gather}

Consider now equations (\ref{case}). It follows from (\ref{cas3})
and (\ref{cas4}) that in our case $g=q$ and so we have the only
system. It is invariant w.r.t. the rotation group and admits
separation of variables in the spherical coordinates. Making changes
(\ref{sv}), (\ref{bes}), using  identities (\ref{a5}), (\ref{a6})
and applying expressions (\ref{sc}) for the second order derivatives
we obtain a system of first order radial equations in the following
form:
\begin{gather}\la{cas}R\frac{\p\Phi}{\p x}=\left(M\frac1{r}-kN+Kr\right)\Phi\end{gather}
where
\begin{gather}\la{not2}\begin{split}&R=
\ri\{\hat S_1,\hat S_2\}=\ri[\hat S_1^2,{\bf S}\cdot{\bf J}], \quad
M=\frac12\hat S_1^2{\bf L}^2-\hat S_2^2+\ri\hat S_1\hat S_2,\\&
\quad N=\{{\bf S}\cdot{\bf J},\tilde \Lambda\},\quad
K=(2q+1)^2-k^2\tilde\Lambda-\hat S_1^2\end{split}\end{gather} and
$\hat S_2=\ri[\hat S_1, {\bf S}\cdot{\bf J}].$ Here $[.,.]$ and
$\{.,.\}$ denote commutator and anticommutator respectively.

 Using their definitions, all matrices (\ref{not2}) can be found in the explicit forms:
 \begin{gather}\la{ma}\begin{split}&R=\mu\left(\begin{matrix}0&0&-1\\0&0&0\\1&0&0
 \end{matrix}\right),
 \quad M=-\mu\left(\begin{matrix} 0&0&j+1\\0&0&0\\j&0&0\end{matrix}\right),\quad N=\frac{\mu}{2j+1}\left(\begin{matrix}2\mu&0&-1\\0&0&0\\-1&0&-2\mu\end{matrix}\right),
 \\& K=((2q+1)^2-1)I+(1-k^2)\hat\Lambda,\quad \tilde\Lambda=\frac1{2j+1}\left(\begin{matrix}j&0&-\mu\\0&0&0\\-\mu&0&j+1
 \end{matrix}\right)\end{split}\end{gather}
 where $I$ is the unit matrix, and
 $\mu=\sqrt{j(j+1)}.$

 It follows from (\ref{cas}), (\ref{ma}) that if $q\neq0$ then $\Phi_0=0$, and equation (\ref{cas}) is reduced to the following form:
 \begin{gather}\left(\frac\p{\p r}-W\right)\tilde\Phi=0\la{sch}\end{gather}
 where $\tilde\Phi=\text{column}(\Phi_-,\Phi_+)$ and $W$ is a matrix superpotential
 \begin{gather}\la{msp}W=\hat M\frac1{r}-k\hat N+\hat Kr\end{gather} with
 \begin{gather*}\hat M= \left(\begin{matrix}j&0\\0&-(j+1)\end{matrix}\right),\quad \hat N=\frac1{2j+1}\left(\begin{matrix}1&2\mu\\2\mu&-1\end{matrix}\right),\\ \\ \hat K=4q(q+1)\left(\begin{matrix}0&-1\\1&0\end{matrix}\right)+\frac{k^2-1}{\mu
 (2j+1)}
 \left(\begin{matrix}-\mu&j+1\\-j&\mu\end{matrix}\right).\end{gather*}

A differential consequence of (\ref{sch})  is the following second-order equation:
\begin{gather}\la{co1}\left(-\frac{\p ^2}{\p r^2}+W^2-W'\right)\hat\Phi=0\end{gather}
where we denote for short $W'=\frac{\p W}{\p r}$.

In order  equations (\ref{sc}) and (\ref{co1}) to be  compatible the matrix $\hat K$ should be nilpotent, i.e., $\hat K^2$=0. This condition generates the following algebraic constrains on parameters $q$ and $k$:
\begin{gather}\la{con3}k^2=(2q+1)^2\end{gather}
or, alternatively, $q=0$.

Let condition (\ref{con3}) is satisfied, then, in accordance with
(\ref{nota}), the admissible energy values for the coupled states
are given by  formula (\ref{E}). The total angular momentum $\bf J$
is a sum of two commuting angular momenta $\hat {\bf q}$  and  $\hat
{\bf g}$ (see (\ref{qg})),  both of which realize irreducible
representation $D(q)$ of algebra so(3). Thus, the admissible values
of $j$ are $2q,2q-1,2q-2,...,0$, and it is possible to represent the
related energy values as
\begin{gather}\la{eva}E=-\frac{m\alpha^2}{2(n+j+1)^2},\quad
n=0,1,2,...\end{gather}

Moreover, the corresponding LRL vector (\ref{GRL}), (\ref{lala}) and
total orbital momentum (\ref{tm}) realize representations $D(0,N)$
of algebra so(4), where $N=n+j+1$.

Since $g=q$, the condition $q=0$ corresponds to the trivial
realization $D(1,0)$ with $N=1$, $j=0$ and $E=-\frac{m\alpha^2}2$.

\subsection{Hamiltonian eigenvectors}
Thus we obtain the spectrum of hamiltonian (\ref{EP}) in the form  (\ref{E}) without solving the corresponding Schr\"odinger equation. To find this spectrum it was sufficient to ask for the compatibility of equations (\ref{resc}) and (\ref{case}).

Let us find the corresponding eigenvectors of hamiltonian (\ref{EP}). Instead of the system of second order equations (\ref{resc}) it is sufficient to solve the first order equations (\ref{sch}) which
are reduced to the following system:
\begin{gather}\la{sys}\begin{split}& \Phi_-' = \frac{j}{r}\Phi_--\frac{k}{2j+1}(\Phi_- +2\mu\Phi_+)-\frac{(k^2-1)r}{(2j+1)\mu}(j\Phi_++\mu \Phi_-),\\&
\Phi_+'=-\frac{j+1}{r}\Phi_++\frac{k}{2j+1}\left(\Phi_+-2\mu\Phi_-\right)+\frac{(k^2-1)r}{\mu(2j+1)}\left((j+1)\Phi_-+\mu\Phi_+\right).\end{split}\end{gather}

This system can be simplified by the following change of dependent variables:
\begin{gather}\la{che}\Phi_-=\frac1{\sqrt{2j+1}}\left(\sqrt{j+1}
\tilde\Phi_-+\sqrt{j}\tilde\Phi_+\right),\quad \Phi_+=\frac1{\sqrt{2j+1}}\left(\sqrt{j+1}
\tilde\Phi_+-\sqrt{j}\tilde\Phi_-\right)\end{gather}
(this transformation is unitary). As a result we obtain:
\begin{gather}\la{sys1}\tilde\Phi_+=-\frac{r}\mu(\tilde\Phi_-'+k\tilde\Phi_-),\\\la{sys2}\tilde\Phi_+'=k\tilde\Phi_+-\frac1{r}\tilde\Phi_+-\frac\mu{r}\tilde\Phi_-+\frac{k^2-1}\mu\tilde\Phi_-.\end{gather}

Substituting (\ref{sys1}) into (\ref{sys2}) we obtain the following equation:
\begin{gather*}-r^2\tilde\Phi_-''-2r\tilde\Phi_-' +(j(j+1)-2kr+r^2)\tilde\Phi_-=0\end{gather*}
This equation  is solved by the following function
\begin{gather}\la{phi1}\tilde\Phi_-=C_{kj}r^{j}\exp(-r){\cal F}(j+1-k,2j+2,2r)\end{gather}
where $\cal F$ is the confluent hypergeometric function and $C_{kj}$ is an integration constant. The corresponding function $\tilde\Phi_+$ is defined by equation (\ref{sys1}) and is easy calculated:
\begin{gather}\la{phi2}\begin{split}&\tilde\Phi_+=-\frac1{\mu}{C_{kj}}
r^{j}\exp{(-r)}\left(((k-1)r+j+1){\cal F}(j+1-k,2j+2,2r)\right.\\&\left.+\frac{j+1-k}{j+1}r {\cal F}(j+2-k,2j+3,2r)\right).\end{split}\end{gather}

Functions (\ref{phi1}) and (\ref{phi2}) are square integrable
provided the first argument of the hypergeometric functions is a
negative integer. Moreover, thanks to the multiplier $j+1-k$ in the
last line of equation (\ref{phi2}) the admissible values of $k$ and
$j$ are: $k=j+1+n,\quad n=0,1,...$. Then formula (\ref{nota}) gives
the   energy values (\ref{eva}) for coupled states, which is in a
good accordance with the discussion presented in the previous
section. The corresponding eigenvectors of hamiltonian
(\ref{hamham}) are given by equations (\ref{bes}) and (\ref{not1})
where, in accordance with (\ref{che}), (\ref{phi1}) and
(\ref{phi2}),
\begin{gather*}\begin{split}&\psi_{j\kappa-1}=C_{jn}\sqrt{j}r^{j+1}\exp(-r)
\left((j+n)(j+1){\cal F}(-n,2j+2,2r)-{n}{\cal F}(1-n,2j+3,2r)
\right),\\
&\psi_{j\kappa+1}=C_{jn}\sqrt{j+1}r^{j}\exp(-r)\left(nr{\cal F}
(1-n,2j+3,2r)\right.\\&\left.-(j+1)(2j+1+(n+j)r){\cal
F}(-n,2j+2,2r)\right),\quad
\psi_{j\kappa0}=0.\end{split}\end{gather*}
\section{The system with spin $\frac32$}
Consider the LRL vector for spin $\frac32$. The corresponding spin
matrices are:
  \begin{gather*}\begin{split}&S_1=\frac12
  \begin{pmatrix}0&\sqrt{3}&0&0\\\sqrt{3}&0&2&0\\0&2&0&\sqrt{3}\\
  0&0&\sqrt{3}&0
  \end{pmatrix},\quad S_2=\frac{\ri}2\begin{pmatrix}0&-\sqrt{3}&0&0\\
  \sqrt{3}&0&-2&0\\0&2&0&-\sqrt{3}\\
  0&0&\sqrt{3}&0
  \end{pmatrix}\end{split}\label{3/2}\end{gather*}
and
\begin{gather*}
S_3=\frac12
  \begin{pmatrix}3&0&0&0\\0&1&0&0\\0&0&-1&0\\
  0&0&0&-3
  \end{pmatrix}.\end{gather*}

 The related hamiltonian (\ref{HS}) takes the following form:
\begin{gather}\la{hamhamham}H=\frac{p^2}{2m}+\frac{\alpha}{x}
\hat\Lambda,\quad \hat\Lambda=\frac43\left(5{\bf S}\cdot{\bf n}-
2({\bf S}\cdot{\bf n})^3\right).\end{gather}

The  eigenvalue problem for this hamiltonian
 in rescaled variables (\ref{change}) is reduced to the standard form
 (\ref{resc}) where $\hat\Lambda$ is the matrix defined in (\ref{hamhamham}).
By construction, this equation admits six integrals of motion which are the
 components of total orbital momentum $\bf J$ and LRL vector
 $\hat {\bf K}$. Thus, like in sections 5 and 6.1, we can search for values of the spectral parameter $k$ algebraically.

The eigenvalue problems for the Casimir operator $C_\pm$ admit the
uniform representation (\ref{case}), (\ref{cas7}) where $s=\frac32$.
Introducing the spherical variables and expanding solutions of the
related equations (\ref{resc}) and (\ref{case}) via spherical
spinors like in (\ref{bes}), we obtain the following systems of
equations for radial functions:
\begin{gather}\la{ba1}\begin{split}&\Phi_1''=\Phi_1+\frac{(2j-1)(2j-3)}{4r^2}\Phi_1-\frac {2k}{\mu r}(\sqrt{3}j\Phi_3-\delta\Phi_4),\\&\Phi_2''=\Phi_2+\frac{(2j+1)(2j+3)}{4r^2}-
\frac{2k(j+1)\sqrt{3}}{\mu r}\Phi_4,\\&\Phi_3''=\Phi_3+\frac{(2j+1)(2j-1)}{4r^2}\Phi_3-\frac{2kj\sqrt{3}}{\mu r}\Phi_1,\\&\Phi_4''=\Phi_4+\frac{(2j+3)(2j+5)}{4\mu r}\Phi_4+\frac{2k}{\mu r}(\delta\Phi_1-(j+1)\sqrt{3}\Phi_2)\end{split}\end{gather}
and
\begin{gather}\begin{split}\la{ba2}&\Phi_1'  =
\frac{\left( 2\,j-1 \right)}{2r} \Phi_1 -\frac {k}{\mu}
\left( \delta \Phi_4+\sqrt{3}\Phi_3\right)
 +r\sum_c R_{1c}\Phi_c,\\&\Phi_2' = -\frac
 { \left( 2\,j+1 \right)}{2r} \Phi_2-\frac{k}\mu\left(
 \delta \Phi_3-\sqrt{3}\Phi_4\right) +r\sum_c R_{2c}\Phi_c,
\\&\Phi_3'  =\frac { \left( 2\,j+1 \right)}{2r} \Phi_3  -\frac{k}\mu\left(\delta \Phi_2+\sqrt{3}\Phi_1\right)
 +r\sum_c R_{3c}\Phi_c,
\\&\Phi_4' =-\frac { \left( 2\,j+3 \right)}{2r} \Phi_4  -\frac{k}\mu\left(\delta \Phi_1-\sqrt{3}\Phi_2\right)
 +r\sum_c R_{4c}\Phi_c\end{split}\end{gather}
where we denote for short
$\psi_{j, \kappa,\lambda}(r)=$ column$(\Phi_1,\Phi_3, \Phi_2, \Phi_4)$, and
\begin{gather}\la{ba4}\begin{split}&R_{11}=-R_{22}=\frac{1-4k^2}{2j\delta},\quad R_{33}=-R_{44}=
\frac{1-4k^2}{2(j+1)\delta}, \\&
R_{12}=\frac1j(12k^2(j+1)-4j\omega_{\pm}-3+7j), \\&
R_{21}=\frac1j(4\omega_{\pm} j-4k^2(7j-3)-3j-3),\\&
R_{34}=\frac1{j+1}(4k^2(7j+10)+3j-4-4\omega_{\pm}(j+1)),\\&
R_{43}=\frac1{j+1}(4\omega_{\pm}(j+1)-12jk^2-7j-10).\end{split}\end{gather}
The remaining matrix entries $R_{nc}\ (n,c=1,2,3,4)$ in (\ref{ba2})
 are equal to zero.

A necessary condition of compatibility  of equations
(\ref{ba1})--(\ref{ba2}) is $R^2=0$ where $R$ is a matrix whose
non-zero entries are given in (\ref{ba4}). This condition (which
appears to be also sufficient) is
 reduced to the following constraints for the
spectral parameter $k$:
\begin{gather}\la{ba5}k^2=\omega_\pm-\frac94\end{gather}
or, alternatively,
\begin{gather}\la{ba6}(3k)^2=\omega_\pm-\frac14.\end{gather}

Using definitions (\ref{cas7}) for $\omega_\pm$ we rewrite
conditions (\ref{ba5}) and (\ref{ba6}) as:
\begin{gather}k=n+\frac32, \quad n=2q+1=1,2,..., g=q+\frac32\la{4}\end{gather}
and
\begin{gather}\la{5}k=\frac13\left(n+\frac12\right),\quad n=2q+1=1,2,...,
g=q+\frac12\end{gather} respectively.

  In accordance with
(\ref{nota}), the corresponding energy levels have  the form
(\ref{bb1}) with \begin{gather}\la{N2}N=n+\frac32={\tilde n}+j,\
j=\frac32, \frac52..., \ \tilde n=1,2,\dots\end{gather} for the case
(\ref{4}) or, alternatively,
\begin{gather}\la{EV2}E=-\frac{9m^2\alpha^2}{2N^2},
\quad N=n+\frac12=\tilde n+j, \quad \tilde n=1,2,...\end{gather} for
the case (\ref{5}). In the latter case $j$ can take any half integer
value including $j=\frac12$.  The  eigenvectors corresponding  to
the presented eigenvalues can be expressed via solutions of the
fourth order scalar ordinary equations, which we obtain by excluding
three out of four dependent variables from the first order system
(\ref{ba2}), see Appendix B for detailed calculations.

Thus there are two possible spectrum branches which are generated by
the compatibility condition for the systems (\ref{ba1}) and
(\ref{ba2}). They can be found algebraically using rather simple
calculations. However, to calculate the corresponding eigenvectors
is not an easy job, see Appendix B.

Using (\ref{N2}) and (\ref{EV2}) it is possible to specify the
irreducible representations of algebra so(4) which are generated by
the total momentum and LRL  vectors. Namely, they are
representations $D(\frac12,N)$ for the case (\ref{N2}) and
$D(\frac32,N)$ for the case (\ref{EV2}).

\section{A physical interpretation}

The  systems discussed above describe particles with spin
interacting with an external field. Since the systems hamiltonians
do not include minimal interaction terms, these particles are
neutral but can have nontrivial dipole or multipole momenta. Let us
consider some representations of the considered systems and discuss
their possible interpretations.

For $s=\frac12$ the corresponding potential (\ref{V}) represents the
dipole interaction with a vector external field $F$ whose components
are:
\begin{gather}F_a=\frac{x_a}{x^2}.\la{ef}\end{gather}

Vector (\ref{ef}) can be interpreted as the electric field strength.
Such a field can be realized experimentally at least on the finite
interval $a < x < b, a > 0$, see, e.g., problem 1018 in \cite{Qum}.

The potential (\ref{lala}) for  $s=1$ can be rewritten as:
\begin{gather}\la{N6}\hat V=\hat Q_{ab}\frac{\p \hat F_a}{\p x_b}\equiv\frac12Q_{ab}\frac{\p \hat F_a}{\p x_b}+\frac16\frac{\p F_a}{\p x_a}
\end{gather}
where $\hat Q_{ab}=\frac12\left(S_aS_b+S_bS_a-\delta_{ab}\right)$,
$Q_{ab}=S_aS_b+S_bS_a-\frac43\delta_{ab}$ is the traceless quadruple
interaction tensor and
\begin{gather}\la{N7}\hat F_a=-\frac{\alpha
x_a}x\end{gather} is a vector of the  external field strength. In
other words, this potentials
 represents a superposition of quadrupole and Darwin interactions of
 spin one
 particle with an external field.

 Alternatively, it is possible to represent potential (\ref{lala})
 as a nonlinear function of field (\ref{ef}):
 \[\hat V=\alpha\frac{{\bf F}^2-({\bf S}\cdot{\bf F})^2}{|{\bf
 F}|}.\]
 Let us recall that a nonlinear generalization of the Pauli interaction
 is also required to obtain a consistent
relativistic description of spin-1 particle interacting with an
external field \cite{becky}.

 The potential for spin $\frac32$ presented in equation
 (\ref{hamhamham}) can be represented as
 \begin{gather} \hat V=
  \frac{2\alpha}9 Q_{abc}\frac{\p^2 \tilde F_a}{\p x_b\p x_c}\label{IHIHI}
  \end{gather}
where
\begin{gather}\la{N8}\tilde F_a=x_a\ln{x},\end{gather}
and
\begin{gather}\la{N9}Q_{abc}=\sum_{P(a,b,c)}\left(S_aS_bS_c-
\frac{7}4S_a\delta_{bc}\right)\end{gather}
 is the octupole
interaction tensor. In equation (\ref{N9}) the summation is imposed
over all possible permutations of indices $a, b$ and $c$.

Alternatively, this potential can be rewritten as:
\[\hat V=\alpha\frac43\left(5{\bf S}\cdot{\bf F}-\frac{2({\bf
S}\cdot{\bf F})^3}{|{\bf F|^2}}\right)\] where $\bf F$ is a vector
whose components are given by equation (\ref{ef}).

A more difficult task is a physical interpretation of vectors
(\ref{N7}) and (\ref{N8}) since the corresponding classical electric
fields needs charge densities which are hardly realized
experimentally. However, all vectors (\ref{ef}), (\ref{N7}) and
(\ref{N8}) solve equations of the axion electrodynamics \cite{Nku1},
\cite{Nku2}, and this fact presents additional interesting
possibilities for their interpretation.

\section{Discussion}
Thus we generalize the Fock  symmetry of the nonrelativistic
Hydrogen atom to the case of arbitrary spin. More exactly, for any
spin $s$ we specify  a QM system which admits two vector integrals
of motion  which are the total orbital momentum and the generalized
LRL vector with arbitrary spin. Moreover, they generate the hidden
symmetry of coupled states of the specified QM systems w.r.t. group
SO(4).

To construct the LRL vector for arbitrary spin we postulate its
generic  form (\ref{GRL}) and find potentials $\hat V$ for which
such vector commutes with the corresponding hamiltonian (\ref{HS}).
The determining equations (\ref{cond1})--(\ref{cond3}) for the
potential can be solved exactly for any $s$. Moreover, their
solution is unique up to a coupling constant.

The representation of the generalized LRL vectors in form
(\ref{GRL}) was chosen by analogy with the LRL for the Hydrogen
atom. However, at least for $s=0$ and $s=\frac12$ this
representation is unique for all generic hamiltonians (\ref{HS})
with a (matrix) potential $\hat V$ depending on  $\bf x$. For a
scalar hamiltonian this fact was proven, e.g., in \cite{evans} where
all second order integrals of motion for 3d QM systems had been
classified. For spin $\frac12$ the uniqueness of the generalized LRL
vector was proven in paper \cite{N2} were all the related
superintegrable systems which are invariant w.r.t. rotation group
and admit second order integrals of motion where specified.

Like in the case of the Hydrogen atom, the generalized LRL vector
can be used to find the spectrum of the corresponding hamiltonian
algebraically. For $s=\frac12, 1$ and $\frac32$ it has been done in
sections 5, 6.2 and 7 correspondingly. However, except the case
$s=\frac12$, the considered superintegrable systems with spin do not
succeed one more hidden symmetry of the Hydrogen atom, i.e., the
shape invariance. Indeed, the lists of shape invariant matrix
potentials presented in \cite{NK1} and \cite{NK2} does not include
effective potentials of equations (\ref{sc}) and (\ref{ba1})
\footnote{These lists ignores potentials with the novel enlarged
shape invariance property discussed, e.g., in \cite{raca}. The
classification of matrix potentials which possess this property is
still an open problem}. As a result the construction of eigenvectors
of the corresponding Hamiltonians is a rather sophisticated problem.
However the extended number of integrals of motion admitted by these
systems makes it possible effectively simplify the calculation of
eigenvectors and solve the system of the first order equations of
the form (\ref{sch}) instead of  of coupled Schr\"odinger equations,
like it was done in sections 6.3 and Appendix 2.

It is interesting to compare the results of the present paper with
ones obtained in \cite{pron2} and \cite{N2} where  2d
superintegrable systems with arbitrary spin which admit two
dimensional analogues of the LRL vector were discussed. As it was
shown in \cite{N2}, the 2d systems are both superintegrable and
supersymmetric. Namely, they admit 2d LRL vectors and are shape
invariant. The latter circumstance makes it possible to find their
exact solutions in exact form for any spin $s$ \cite{N2}.

In the 3d case considered in the present paper the situation is more
complicated. The systems with spins 0 and $\frac12$ are shape
invariant, but starting with spin 1 this property is loosed, and the
procedure of  constructing  of exact solutions is rather
sophisticated.

Comparing the system with spin 1, whose potential and hamiltonian
are given by equations (\ref{lala}) and (\ref{hamham}), with the
corresponding 2d system discussed in \cite{N2} (see equation (4.6)
there) we conclude, that their potentials are qualitatively
different. Indeed, potential (\ref{lala}) is a product of the
Coulomb potential and matrix $\hat\Lambda$ (\ref{rescaca}) which is
a projector satisfying $\hat\Lambda^2=\hat\Lambda$. The
corresponding matrix in the 2d model is not a projector and its
square is proportional to the unit matrix. Nevertheless, up to
notations of the coupling constants and possible values of the
summation indices, representation (\ref{N6}) is valid for both the
3d and 2d cases.

The same is true for systems with spin $\frac32$. Again the 3d and
2d matrix potentials are qualitatively different since $\hat\Lambda$
in (\ref{hamhamham}) is a matrix inverse to ${\bf S}\cdot{\bf n}$
while the corresponding matrix in \cite{N2} is proportional to the
square root of the unit matrix. However, representation
(\ref{IHIHI}) is valid for both the 3d and 2d systems. In addition,
in both cases there are two branches of energy spectrum given by
equations (\ref{bb1}), (\ref{N2}) and (\ref{EV2}) (but in 2d case
$j$ is replaced by eigenvalues of $J_3$).

Thus we extend the results of papers \cite{pron2}, \cite{N2} and
\cite{N1} to the case of 3d systems with arbitrary spin. The next
task is to find relativistic counterparts of the discussed systems
like it was done in \cite{N1} and \cite{Nme} for $s=\frac12$. We
plane to do it using the approach developed in \cite{NN1} and
\cite{NN2}.

There are also other challenges connected with the presented
results. Among them is a search for LRL vectors with spin in multi
dimensional and curved spaces,  discussion of superintegrable
superpartners of the presented systems and various generalizations
with quadratic algebras of higher symmetries. For a generalized
MICZ-Kepler system which looses the symmetry w.r.t. algebra so(4)
but is integrable thanks to the quadratic symmetry algebra see
\cite{Mard} and \cite{Marq}.

\appendix
\section{Spherical harmonics and invariant matrices}
\renewcommand{\theequation}{A\arabic{equation}} \setcounter{equation}{0}
Here a definition of spherical harmonics and explicit forms of
rotationally invariant matrices, used in the main text, are
presented. They can be found, e.g., in \cite{bied}, but we use the
notations presented in \cite{FN}.

 Spherical harmonics are eigenvectors of the commuting operators $ J_3$ , ${\bf J}^2,\ {\bf S}^2$ and operator ${\bf L}^2$.  By definition they satisfy the following equations:
\begin{gather}\la{EV1}\begin{split}&{\bf J}^2\Omega^s_{j,\kappa,\lambda}=
j(j+1)\Omega^s_{j,\kappa,\lambda},\\&J_3\Omega^s_{j,\kappa,\lambda}=
\kappa\Omega^s_{j,\kappa,\lambda},\\&{\bf L}^2\Omega_{j,\kappa,\lambda}=
(j-\lambda)(j-\lambda+1)\Omega^s_{j,\kappa,\lambda}\\&{\bf S}^2\Omega^s_{j,\kappa,\lambda}=s(s+1)\Omega^s_{j,\kappa,\lambda}.
\end{split}\end{gather}
Here $s$ are  integers or  half integers labeling  irreducible representations of the rotation group,
\begin{gather}\la{a2}j=0,\ 1,\ \dots, \ \kappa=-j,\ -j+1,\ \dots\ j, \ \lambda=-s,-s+1,...,-s+2\min(s,j).\end{gather}

The spherical harmonic can be represented as a column whose $\mu$-th component is given by the following equation:
\begin{gather}\la{a3}(\Omega^s_{j, \kappa, \lambda})_\mu=C_{j-\lambda, \kappa-\mu, s, \mu}Y_{j-\lambda, \kappa-\mu},\quad \mu=1,2,...,2\lambda+1\end{gather}
where $Y_{..}$ are spherical functions and $C_{....}$ are Wigner coefficients. In addition, the following normalization condition is satisfied:
$$\int \Omega^{s\dag}_{j, \kappa, \lambda}\Omega^{s\dag}_{j', \kappa', \lambda'}d\omega=\delta_{jj'}\delta_{\kappa\kappa'}\delta_{\lambda\lambda'}$$
where $\delta_{jj'}$ is the Kronecker symbol and $d\omega=\sin \theta d\theta d\phi$.

In the spherical harmonics basis the matrix ${\bf S}\cdot{\bf n}$ is reduced to the following form:
\begin{gather}{\bf S}\cdot{\bf n}\Omega^{s\dag}_{j, \kappa, \lambda}=-\frac12\left(A^{sj}_{j-\lambda-1}\Omega^{s\dag}_{j\ \kappa\ \lambda-1}+A^{sj}_{j-\lambda+1}\Omega^{s\dag}_{j,\kappa, \lambda+1}\right)\la{a4}\end{gather}
where
\begin{gather*}A^{sj}_\mu=\left(\frac{\mu(2j+1-\mu)(2s+1-\mu)(2j+2s-2-\mu)}
{(2j+2s-2\mu+1)(2j+2s-2\mu+3)}\right)^\frac12.\end{gather*}

The matrix ${\bf J}^2$ is diagonal. Its entries are given by the
first equation (\ref{EV1}). The matrix ${\bf S}\cdot{\bf J}$ is
diagonal also, its entries are easy calculated using the identity
\begin{gather}\la{a5}{\bf S}\cdot{\bf J}=\frac12\left(j(j+1)+s(s+1)-{\bf L}^2\right).\end{gather}

We used one more rotationally invariant matrix, i.e., ${\bf S}\cdot{\bf p}$, whose entries are differential operators. When acting on functions expanded via spherical harmonics basis like (\ref{bes}) it takes the following form:
\begin{gather}{\bf S}\cdot{\bf p}=-\imath {\bf S}\cdot{\bf n}\frac{\p}{\p x}+\frac{\imath}{x}[{\bf S}\cdot{\bf n},{\bf S}\cdot{\bf L}].\la{a6}\end{gather}

Just definition (\ref{a3}) for $s=\frac12$ and the momentum
representation was used in equation (\ref{OOmu}).

\section{ Some calculation details for spin $\frac32$}
\renewcommand{\theequation}{B\arabic{equation}} \setcounter{equation}{0}
In the case of spin $\frac32$ we are supposed to solve three systems
of equations, i.e., (\ref{ba1}) and (\ref{ba2}).

 If condition (\ref{4}) is satisfied,  system (\ref{ba2})
takes the following form:
\begin{gather}\begin{split}\la{b3}&\Phi_1'  =\frac{\left( 2\,j-1 \right)}{2r} \Phi_1 -\frac {k}{\mu} \left( \delta \Phi_4+\sqrt{3}\Phi_3\right)
 -\frac{(4k^2-1)r}{2j}\left(\Phi_1 +\frac{(2j-1)\sqrt{3}}{\delta}\Phi_2\right),\\&\Phi_2' = -\frac { \left( 2\,j+1 \right)}{2r} \Phi_2-\frac{k}\mu\left(\delta \Phi_3-\sqrt{3}\Phi_4\right)+\frac{(4k^2-1)r}{2j}\left( \frac{2j+3}{\sqrt{3}\delta}\Phi_1+\Phi_2\right),\end{split}
\\\la{b7}\begin{split}&\Phi_3'  =\frac { \left( 2\,j+1 \right)}{2r} \Phi_3  -\frac{k}\mu\left(\delta \Phi_2+\sqrt{3}\Phi_1\right)
-\frac{(4k^2-1)r}{2(j+1)}\left(\Phi_3
+\frac{2j-1}{\sqrt{3}\delta}\Phi_4\right),
\\&\Phi_4' =-\frac { \left( 2\,j+3 \right)}{2r} \Phi_4  -\frac{k}\mu\left(\delta \Phi_1-\sqrt{3}\Phi_2\right)
+\frac{(4k^2-1)r}{2(j+1)}\left(
\frac{(2j+3)\sqrt{3}}{\delta}\Phi_3+\Phi_4\right).\end{split}\end{gather}

On the other hand, for the case (\ref{5}) we obtain:
\begin{gather}\la{b8}\begin{split}&\Phi_1'  =\frac{\left( 2\,j-1 \right)}{2r} \Phi_1 -\frac {k}{\mu} \left( \delta \Phi_4+\sqrt{3}\Phi_3\right)
 -\frac{(4k^2-1)r}{2j}\left(\Phi_1 -\frac{2j+3}{\sqrt{3}\delta}\Phi_2\right),\\&\Phi_2' =-\frac { \left( 2\,j+1 \right)}{2r} \Phi_2-\frac{k}\mu\left(\delta \Phi_3-\sqrt{3}\Phi_4\right)-\frac{(4k^2-1)r}{2j}\left( \frac{(2j-1)\sqrt{3}}{\delta}\Phi_1-\Phi_2\right),\end{split}
\\\begin{split}&\Phi_3'  =\frac { \left( 2\,j+1 \right)}{2r} \Phi_3  -\frac{k}\mu\left(\delta \Phi_2+\sqrt{3}\Phi_1\right)
-\frac{(4k^2-1)r}{2(j+1)}\left(\Phi_3
-\frac{(2j+3)\sqrt{3}}{\delta}\Phi_4\right),
\\&\Phi_4' =-\frac { \left( 2\,j+3 \right)}{2r} \Phi_4  -\frac{k}\mu\left(\delta \Phi_1-\sqrt{3}\Phi_2\right)
-\frac{(4k^2-1)r}{2(j+1)}\left(
\frac{2j-1}{\sqrt{3}\delta}\Phi_3-\Phi_4\right).\end{split}\la{b4}\end{gather}

Substituting (\ref{b3}) and (\ref{b7}) or (\ref{b8}) and (\ref{b4})
into (\ref{ba1}) we turn the latter equation to identity.  It means
that instead of the second order system (\ref{ba1}) with an unknown
spectral parameter $k$ we can solve the first order system
(\ref{b3}), (\ref{b7}) or (\ref{b8}), (\ref{b4}) with the specified
$k$ given by equation (\ref{4}) or (\ref{5}). Let us consider these
systems consequently.

Solving (\ref{b3}) for $\Phi_3$ and $\Phi_4$ we obtain
\begin{gather}\la{b9}\begin{split}&\Phi_3=\frac1{4k\mu}\left((4k^2-1)r
\left(a^{-1}\Phi_2+
\frac1{\sqrt{3}}\Phi_1\right)-\sqrt{3}\left(\Phi_1'-\frac{2j-1}{2r}\Phi_1\right)
\right.\\&\left.-\delta\left(\Phi_2'+\frac{2j+1}{2r}\Phi_2\right)\right),\\&
\Phi_4=-\frac1{4k\mu}\left((4k^2-1)r\left(a\Phi_1+\sqrt{3}\Phi_2\right)
+
\sqrt{3}\left(\Phi_2'+\frac{2j+1}{2r}\Phi_2\right)\right.\\&\left.-\delta\left(\Phi_1'-
\frac{2j-1}{2r}\Phi_1\right)\right)
\end{split}\end{gather}
where $a=\sqrt{\frac{2j+3}{2j-1}}$.

Substituting these expressions into (\ref{b7}), we obtain a system
of second order equations for two unknowns $\Phi_1$ and $\Phi_2$:
\begin{gather}\la{b101}\begin{split}&\Phi_1''=\frac1{2jr}\left(({3-4j})\Phi_1'+
\delta\sqrt{3}\Phi_2'\right)+\frac1{4jr^2}\left(
\delta^2(j-1)\Phi_1+\sqrt{3}\delta(2j+1)\Phi_2\right)\\&
+\frac1{2j}\left(({2j-12k^2+3})\Phi_1-
{3a^{-1}\sqrt{3}(1-4k^2)}\Phi_2\right),\\&\Phi_2''=\frac1{2jr}
\left(\sqrt{3}\delta\Phi_1'-3\Phi_2'\right)+\frac1{4jr^2}\left((2j+1)
(2j^2+3j-3)\Phi_2
-(2j-1)\sqrt{3}\delta\Phi_1\right)\\&+\frac1{2j}\left(({2j-3+12k^2})\Phi_2+
{a\sqrt{3}(4k^2-1)}\Phi_1\right).\end{split}\end{gather}

The change
\begin{gather}\la{b12}\begin{split}&\Phi_1=\sqrt{r}\left
(a\hat\Phi_1+ \frac1{r^2} \hat\Phi_2\right),\quad\Phi_2=\sqrt{r}
\left(\sqrt{3}\hat\Phi_1- \frac{a}{\sqrt{3}r^2}
\hat\Phi_2\right)\end{split}\end{gather} reduces (\ref{b101}) to a
more simple form which does not contain  the first order
derivatives:
\begin{gather}\la{b14}\hat\Phi_1''=\frac1{4r^2}\left((4j^2+6j-1)\hat\Phi_1+
3\delta\hat\Phi_2\right)+\hat\Phi_1,\\\la{b15}\hat\Phi_2''=
\frac1{4r^2}\left((4j^2+2j-3)\hat\Phi_2-{3}\delta
\hat\Phi_1\right)+\frac{12\left(4k^2-1\right)}\delta\hat\Phi_1+\hat\Phi_2.\end{gather}

Solving (\ref{b14}) for $\hat\Phi_2$ and substituting the solution
into (\ref{b15}) we obtain:
\begin{gather}\la{b16}\hat\Phi_2=\frac1{3\delta}
\left({4r^2} (\hat\Phi_1''-\hat\Phi_1)-(4j^2+6j-1)\hat\Phi_1\right)
\end{gather}
and
\begin{gather}\la{FO}\begin{split}&r^4\hat\Phi_1^{IV}+4r^3\hat\Phi_1'''-
(2r^2+2\mu^2-3)r^2\hat\Phi_1'' -4r^3\hat\Phi_1'
\\&+\left(r^4-2(18k^2-\mu^2-3)r^2+\mu^4+\mu^2-\frac32\right)
{\hat\Phi_1}=0.\end{split}\end{gather} Here $\mu$ and $k$ are
parameters given by equations (\ref{mu}) and (\ref{4}).

Thus to find solutions of equation (\ref{ba1}) with the additional
condition (\ref{case}) it is sufficient to solve (\ref{FO}) which is
an ordinary differential equation of fourth order. Then using
definitions (\ref{b9}), (\ref{b9}) and (\ref{b16}) it is possible to
reconstruct all components of eigenvectors of hamiltonian
(\ref{hamham}). Moreover, we are interested in such solutions of
(\ref{FO}) which generate square integrable components given by
(\ref{b9}), (\ref{b9}) and (\ref{b16}).

Equation (\ref{FO}) is much more simple than the initial system of
four second order equations (\ref{ba1}) where both the parameter $k$
and eigenvectors $\Psi$ are unknowns. It seems to be impossible to
find its solutions explicitly. However, it is possible to evaluate
them as formal series whose coefficients satisfy clear recurrence
relations.

The qualitatively analysis of (\ref{FO}) makes it possible to prove
the existence of good solutions at neighborhood of the critical
point $r=0$ and for  $r\to\infty$. Indeed, for infinitely small and
infinitely large $r$ this equation can be reduced to the following
forms:
\begin{gather}\la{FO1}\begin{split}&r^4\hat\Phi_1^{IV}+4r^3\hat\Phi_1'''-
(2\mu^2-3)r^2\hat\Phi_1'' -4r^3\hat\Phi_1'
+\left(\mu^4+\mu^2-\frac32\right)
{\hat\Phi_1}=0.\end{split}\end{gather} and
\begin{gather}\la{FO2}\begin{split}&r\hat\Phi_1^{IV}+4\hat\Phi_1'''-
4r\hat\Phi_1'' -4\hat\Phi_1' +r
{\hat\Phi_1}=0.\end{split}\end{gather} correspondingly. Regular (and
vanishing) at $x=0$ solutions of (\ref{FO1}) are:
\begin{gather}\la{AS}\begin{split}&
\hat\Phi_1=C_1r^{\frac{1+\varkappa_-}2}{\cal
F}\left(\frac14+\frac{\varkappa_-}2,\left[1+\frac{\varkappa_-}2,1+
\frac{\varkappa_--\varkappa_+}4,1+
\frac{\varkappa_-+\varkappa_+}4\right],\frac{r^2}2\right)\\&+C_2r^{\frac{1+\varkappa_+}2}{\cal
F}\left(\frac14+\frac{\varkappa_+}2,\left[1+\frac{\varkappa_+}2,1+
\frac{\varkappa_-+\varkappa_+}4,1+
\frac{\varkappa_+-\varkappa_-}4\right],\frac{r^2}2\right)\end{split}\end{gather}
where ${\cal F}(.,.,.)$ are  hypergeometric functions, $C_1$ and
$C_2$ are arbitrary constants, and
\[\varkappa_\pm=\sqrt{4\mu^2-1\pm\sqrt{7-8\mu^2}}.\]

Equation (\ref{FO2}) also has "good" (i.e., square integrable and
vanishing at $r\to\infty$) solutions given by the following
formulae:
\[\hat \Phi_1=C_1\exp(r)+C_2\frac{\exp(r)}r.\]
In other words, equation (\ref{FO}) has good asymptotic solutions in
critical points $r=0$ and $r\to\infty$.

Analogously, starting with (\ref{b8}) and (\ref{b4}), we can solve
the first pair of equations for $\Phi_3$ and $\Phi_4$ and obtain:
\begin{gather}\la{b21}\begin{split}&\Phi_3=\frac1{4k\mu}\left((4k^2-1)r
\left(a\Phi_2-
\sqrt{3}\Phi_1\right)-\sqrt{3}\left(\Phi_1'-\frac{2j-1}{2r}\Phi_1\right)
\right.\\&\left.-\delta\left(\Phi_2'+\frac{2j+1}{2r}\Phi_2\right)\right),\\&
\Phi_4=-\frac1{4k\mu}\left((4k^2-1)r\left(a^{-1}\Phi_1-\frac1{\sqrt{3}}
\Phi_2\right) +
\sqrt{3}\left(\Phi_2'+\frac{2j+1}{2r}\Phi_2\right)\right.\\&\left.-\delta\left(\Phi_1'-
\frac{2j-1}{2r}\Phi_1\right)\right).
\end{split}\end{gather}
Substituting that into (\ref{b4}) one obtains the following second
order system:
\begin{gather}\la{b22}\begin{split}&\Phi_1''=\frac1{2jr}\left(({3-4j})\Phi_1'+
\delta\sqrt{3}\Phi_2'\right)+\frac1{4jr^2}\left(
\delta^2(j-1)\Phi_1+\sqrt{3}\delta(2j+1)\Phi_2\right)\\&
+\frac1{2j}\left(({2j+4k^2-1})\Phi_1+
{\frac{a}{\sqrt{3}}(1-4k^2)}\Phi_2\right),\\&\Phi_2''=\frac1{2jr}
\left(\sqrt{3}\delta\Phi_1'-3\Phi_2'\right)+\frac1{4jr^2}\left((2j+1)
(2j^2+3j-3)\Phi_2
-(2j-1)\sqrt{3}\delta\Phi_1\right)\\&+\frac1{2j}\left(({2j+1-4k^2})\Phi_2+
{a^{-1}\sqrt{3}(4k^2-1)}\Phi_1\right).\end{split}\end{gather} The
change (\ref{b12}) reduces this system to the following form:
\begin{gather}\la{b23}\hat\Phi_1''=\frac1{4r^2}\left((4j^2+6j-1)\hat\Phi_1+
3\delta\hat\Phi_2\right)+\hat\Phi_1-\frac{4\left(4k^2-1\right)}
\delta\hat\Phi_1,\\\la{b24}\hat\Phi_2''=
\frac1{4r^2}\left((4j^2+2j-3)\hat\Phi_2-{3}\delta
\hat\Phi_1\right)+\hat\Phi_2.\end{gather}

Solving (\ref{b24}) for $\hat\Phi_1$ we obtain:
\begin{gather}\la{b25}\hat\Phi_1=\frac1{3\delta}\left(4r^2(\hat\Phi_2-
\hat\Phi_2'')+(4j^2+2j-3)\hat\Phi_2\right).\end{gather} Then
substituting (\ref{b25}) into (\ref{b23}) we obtain the fourth order
equation for $\hat\Phi_2$:
\begin{gather}\la{b26}\begin{split}&r^4\hat\Phi_2^{IV}+4r^3\hat\Phi_2'''-
(2r^2+2\mu^2-3)r^2\hat\Phi_2'' -4r^3\hat\Phi_2'
\\&+\left(r^4-2(2k^2-\mu^2+1)r^2+\mu^4+\mu^2-\frac32\right)
{\hat\Phi_2}=0.\end{split}\end{gather}

Equation (\ref{b26}) differs from (\ref{FO}) only by the coefficient
for the term $r^2\hat\Phi$. In the critical point $r=0$ and for
$r\to\infty$ this coefficient is not essential.

Thus the general solution of system (\ref{b3}), (\ref{b7}) can be
expressed via solutions of the fourth order equation (\ref{b26})
using ansatzs (\ref{b25}), (\ref{b7}) and (\ref{b21}).

Consider separately the special case $j=\frac12$. In this case there
are only two possible values of  $\lambda$ in (\ref{a2}) and we come
to the following $2\times2$ matrices in the spherical harmonics
basis:
\begin{gather}\la{memor}\begin{split}&{\bf S}\cdot{\bf n}=\frac12\sigma_1,\quad \hat \Lambda=3\sigma_1,\quad\hat L^2=2(2+\sigma_3),\\& {\bf S}\cdot{\bf J}=\frac14-\sigma_3,\quad{\bf S}\cdot\nabla=-\frac12\left(\sigma_1\frac\p{\p r}+\ri\sigma_2\frac1r\right)\end{split}\end{gather}
Substituting all that into equations (\ref{case}) and  (\ref{sc}),
(\ref{hamhamham}) we obtain the algebraic condition (\ref{5}) and
the following system of second-order radial equations:
\begin{gather}\begin{split}&\Phi_1''-\Phi_1-\frac6{r^2}\Phi_1+\frac{2\tilde k}r\Phi_2=0,\\&
\Phi_2''-\Phi_2-\frac2{r^2}\Phi_1+\frac{2\tilde
k}r\Phi_1=0\end{split}\la{b27}\end{gather} where $\tilde
k=3k=n+\frac12, n=1,2,...$

Up to scaling  of independent variables the system (\ref{b27})
coincides with equation (13) for $j=\frac32$ of paper \cite{N1},
where its solutions can be found. The corresponding eigenvalues are
given by equation (\ref{EV2}).

\end{document}